\documentclass[a4paper,12pt]{article}

\begin{document}
\topmargin 0pt \oddsidemargin 0mm

\renewcommand{\thefootnote}{\fnsymbol{footnote}}

\begin{titlepage}

\vspace{5mm}
\begin{center}
{\Large \bf  Hawking Radiation of Apparent Horizon in a FRW Universe
} \vspace{32mm}

{\large Rong-Gen Cai$^{a,}$\footnote{e-mail address:
cairg@itp.ac.cn}, Li-Ming Cao$^{b,}$\footnote{e-mail address:
caolm@apctp.org}, Ya-Peng Hu$^{a,c}$\footnote{e-mail address:
yapenghu@itp.ac.cn}}

\vspace{10mm} {\em $^a$ Institute of Theoretical Physics, Chinese
Academy of Sciences,\\
 P.O. Box 2735, Beijing 100190, China \\
 $^{b}$ Asia Pacific Center for
Theoretical Physics, Pohang,
 Gyeongbuk 790-784, Korea \\
  $^c$ Graduate School of the Chinese Academy of Sciences, Beijing 100039, China}

\end{center}

\vspace{10mm} \centerline{{\bf{Abstract}}}
 \vspace{5mm}
Hawking radiation is an important quantum phenomenon of black
hole, which is closely related to the existence of event horizon
of black hole. The cosmological event horizon of de Sitter space
is also of the Hawking radiation with thermal spectrum. By use of
the tunneling approach, we show that there is indeed a Hawking
radiation with temperature, $T=1/2\pi \tilde r_A$, for locally
defined apparent horizon of a Friedmann-Robertson-Walker universe
with any spatial curvature, where $\tilde r_A$ is the apparent
horizon radius. Thus we fill in the gap existing in the literature
investigating the relation between the first law of thermodynamics
and Friedmann equations, there the apparent horizon is assumed to
have such a temperature without any proof. In addition, we stress
the implication of the Hawking temperature associated with the
apparent horizon.
\end{titlepage}

\newpage
\renewcommand{\thefootnote}{\arabic{footnote}}
\setcounter{footnote}{0} \setcounter{page}{2}


{\it Introduction:} A remarkable feature of black hole is the
existence of event horizon. The event horizon can be viewed as
boundary of black hole. It is the existence of event horizon so
that Hawking~\cite{Haw} found that black hole behaves like a black
body, emitting thermal radiation, with a temperature proportional
to its surface gravity on the event horizon and  Bekenstein
argued~\cite{Bek} that there is an entropy proportional to its
horizon area for a black hole. The Hawking temperature and horizon
entropy together with the black hole mass obey the first law of
black hole thermodynamics~\cite{firstlaw}. Hawking radiation and
black hole entropy help us to deepen our understanding on
properties of quantum gravity, although a completely
self-consistent quantum theory of gravity is still not yet
available so far. A seminal work relevant to Hawking radiation of
black hole is done by Unruh~\cite{Unruh}, who found that an
uniformly accelerating observer with acceleration $a$ in
Minkowskian spacetime can detect a thermal spectrum with
temperature $T=a/2\pi$. Throughout the paper we use the geometric
unit where $c=\hbar=1$, but we explicitly write down the
gravitational constant $G$. The Unruh radiation is closely related
to the existence of Rindler causal horizon for the observer. In
addition,  Gibbons and Hawking in 1977 showed~\cite{GH} that very
similar to black hole horizon, there is also a Hawking radiation
with temperature $T=1/2\pi l$ associated with cosmological event
horizon in de Sitter space, here $l$ is the horizon radius of de
Sitter space.  The existence of the cosmological event horizon is
particularly clear in the static coordinates of de Sitter space,
which is
\begin{equation}
\label{ineq1}
 ds^2= -\left(1-\frac{r^2}{l^2}\right)dt^2
+\left(1-\frac{r^2}{l^2}\right)^{-1}dr^2 +r^2 d\Omega_2^2.
\end{equation}
The cosmological event horizon locates at $r=l$. The Hawking
temperature $T=1/2\pi l$ is measured by a comoving observer at
$r=0$. Note that event horizon is a global concept for a
spacetime. Therefore, locally it is not known whether there is an
event horizon associated with a certain dynamical (time-dependent)
spacetime at some time. Thus this causes the difficulty to discuss
Hawking radiation for a dynamic black hole. More recently Hawyard
{\it et al.}~\cite{Hay} have attacked this issue. Using
Hamilton-Jacobi equation of particles to discuss Hawking radiation
of stationary black hole, see also~\cite{Padd}.

Assuming there is a proportionality between entropy and horizon
area, Jacobson~\cite{Jac} derived the Einstein field equation by
using the fundamental Clausius relation, $\delta Q= TdS$,
connecting heat, temperature and entropy. The key idea is to
demand that this relation holds for all the local Rindler causal
horizon  through each spacetime point, with $\delta Q$ and $T$
interpreted as the energy flux and Unruh temperature seen by an
accelerated observer just inside the horizon. In this way,
Einstein field equation is nothing, but an equation of state of
spacetime. Applying this idea to $f(R)$ theory~\cite{Jac1,AC1,AC}
and scalar-tensor theory~\cite{AC1,CC1}, it turns out that a
nonequilibrium thermodynamic setup has to be employed. For another
viewpoint, see \cite{ES,WuYZ}.

It is quite interesting to note that assuming the apparent horizon
of a Friedmann-Robertson-Walker (FRW) universe has temperature $T$
and entropy $S$ satisfying
\begin{equation}
\label{eq1}
 T=\frac{1}{2\pi \tilde r_A}, \ \ \ S= \frac{A}{4G},
 \end{equation}
 where $\tilde r_A$ is the radius of the apparent horizon and $A$ is
 the area of the apparent horizon,  one is able to
derive Friedmann equations of the FRW universe with any spatial
curvature by applying the Clausius relation to apparent
horizon~\cite{CK}. This works not only in Einstein gravitational
theory, but also in Gauss-Bonnet and Lovelock gravity theories.
Here a key ingredient is to replace the entropy area formula in
Einstein theory by using entropy expressions of black hole horizon
in those higher order curvature theories. For related discussions
see also \cite{FK}. These results should closely relate to the
fact that Einstein field equation can be rewritten as an unified
first law~\cite{Hayward3}. Indeed, at the apparent horizon of FRW
universe, the first Friedmann equation can be cast to a universal
form like the first law of thermodynamics~\cite{CC1,AC2}. Note
that in those works the apparent horizon is also assumed to have a
temperature $T=\kappa/2\pi$ without any proof, here $\kappa$ is
the surface gravity of apparent horizon.  This form also holds in
RSII brane world scenario, warped DGP model and even more
complicated case with a Gauss-Bonnet term in bulk~\cite{CC2}.
Based on this, one is able to find the relation between entropy
expression of apparent horizon and horizon geometry in brane world
scenarios. These results have been summarized in~\cite{Cai}. For
further discussions in this direction see~\cite{Ge}. On the other
hand, there also exist some studies in the relation between
Einstein field equation and first law of thermodynamics in the
setup of black hole spacetime~\cite{Pap}.

However, a crucial problem remains in those investigations
relating Friedmann equation to the first law of thermodynamics.
That is, is there indeed a Hawking radiation with temperature
$T=1/2\pi \tilde {r}_A$ associated with the apparent horizon of
the FRW universe since in those studies it is assumed without
strict proof? Which observer sees the Hawking temperature? In the
present paper we are going to fill in this gap. By applying the
Hamilton-Jacobi method~\cite{Padd} and Parikh-Wilczek
approach~\cite{PW}, which are initially designed to study the
Hawking radiation of stationary black hole as a tunneling process
of particle, we show that as in the case of black hole horizon,
there is indeed a thermal radiation for the apparent horizon of
FRW universe.

{\it Hamilton-Jacobi Method:} Let us start with the FRW metric
\begin{equation}
\label{eq2}
 ds^2 = -dt^2 +a^2(t)\left(\frac{dr^2}{1-kr^2} +r^2
d\Omega_2^2\right),
\end{equation}
where $t$ is the cosmic time, $r$ is the comoving coordinate, $a$
is the scale factor, $d\Omega_2^2$ denotes the line element of a
$2$-dimensional sphere with unit radius, $k=1$, $0$ and $-1$
represent a closed, flat and open FRW universe, respectively.
Define $\tilde r = ar$, the metric (\ref{eq2}) can be rewritten as
$ ds^2 = h_{ab} dx^adx^b + \tilde r^2 d\Omega_n^2, $ where $x^a=
(t, r)$, $h_{ab}= {\rm diag}(-1, a^2/(1-kr^2))$. Without the whole
evolution history of the universe, one cannot know whether there
is a cosmological event horizon. However, apparent horizon always
exists in the FRW universe since it is a local quantity of
spacetime. By definition, $h^{ab}\partial_a\tilde r
\partial_b\tilde r=0$, we can know the location of the apparent
horizon in the FRW universe, $\tilde r =\tilde r_A \equiv
1/\sqrt{H^2+k/a^2}$. Here $H= \dot a/a$ is the Hubble parameter.
When $k=0$, the apparent horizon is just the Hubble horizon. We
see that for a de Sitter space where $H$ is a constant and
cosmological event horizon radius is $H^{-1}$, the apparent
horizon and cosmological event horizon coincide with each other
only in the case of $k=0$.

In order to discuss the tunneling of particle, it turns out
convenient to use the coordinates $(t, \tilde r)$. In that case,
the metric (\ref{eq2}) can be rewritten as
\begin{equation}
\label{eq3}
 ds^2=-\frac{1-\tilde r^2/\tilde r_A^2}{1-k \tilde r^2/a^2}dt^2
 -\frac{2 H\tilde r}{1-k\tilde r^2/a^2}dt d\tilde r +
 \frac{1}{1-k\tilde r^2/a^2}d\tilde r^2 +\tilde r^2 d\Omega_2^2.
 \end{equation}
We note that when $k=0$, the metric is quite similar to the
Painlev\'e-de Sitter metric for the de Sitter space~\cite{Parikh},
where $\tilde r_A=H^{-1}=l$ is a constant.  For the  metric
(\ref{eq3}), the corresponding Kodama vector~\cite{Kodama} is
\begin{equation}
\label{eq5}
 K^{a}\equiv -\epsilon ^{ab}\nabla _{b}\tilde
r=\sqrt{1-k\tilde r^{2}/a^{2}}(\partial /\partial t)^{a}
\end{equation}%
where $\epsilon _{ab}=\frac{1}{\sqrt{1-k\tilde
r^{2}/a^{2}}}(dt)_{a}\wedge (d\tilde r)_{b}$. Thus one has
$K_aK^a=-(1-\tilde r^2/\tilde r_A^2)$. Therefore the Kodama vector
is time-like, null and space-like as $\tilde r <\tilde r_A$,
$\tilde r =\tilde r_A$ and $\tilde r > \tilde r_A$, respectively.
Note that the Kodama vector is very similar to the Killing vector
$(\partial/\partial t)^a$ in the de Sitter space (\ref{ineq1}),
the latter is time-like, null and space-like when $r<l$, $r=l$ and
$r>l$, respectively. Note that the existence of the Kodama vector
will play a crucial role in our discussion.

Before preceding, let us stress some differences between
stationary black hole spacetime and time-dependent dynamical FRW
spacetime. For stationary black hole spacetime, one can define a
time-like Killing vector. By the time-like Killing vector, one is
able to obtain a conserved mass (energy) associated with the
stationary black hole spacetime. Also one can define a conserved
energy of a particle moving in the stationary black hole
spacetime. On the other hand, there is no time-like Killing vector
in the dynamical FRW spacetime. But the Kodama vector defined in
(\ref{eq5}) could play a similar role in the FRW spacetime as the
time-like Killing vector does in the stationary black hole
spacetime. By the Kodama vector, one can define a conserved
quantity~\cite{Ha}, Misner-Sharp energy~\cite{MS}, for the FRW
spacetime, which plays a crucial role in investigating the
relation between the first law and Friedmann
equations~\cite{CC1,AC2}. By the time-like Kodama vector inside
the apparent horizon, one can also therefore define a conserved
energy of a particle moving in the FRW spacetime, very like the
case for the time-like Killing vector in the stationary black hole
spacetime. This point is crucial in the following discussion.

Following the discussion for a dynamical black hole~\cite{Hay}, we
consider a particle with mass $m$ radially moving in the
background (\ref{eq3}). The Hamilton-Jacobi equation is
\begin{equation}
g^{\mu \nu }\partial _{\mu }\mathbf{S}\partial _{\nu
}\mathbf{S}+m^{2}=0.
\end{equation}
By use of the Kodama vector (\ref{eq5}), one therefore can define
the energy $\omega$ and radial momentum  $k_{\tilde r}$ associated
with the particle
\begin{equation}
\label{eq7}
 \omega =-K^{a}\partial _{a}\mathbf{S}=-\sqrt{1-k\tilde
r^{2}/a^{2}}\partial _{t}\mathbf{S},\ \ \ k_{\tilde r}=(\partial
/\partial \tilde r)^{a}\partial _{a}\mathbf{S}=\partial _{\tilde
r}\mathbf{S}.
\end{equation}
Thus the action $\mathbf{S}$ can be written as
\begin{equation}
\mathbf{S}=-\int \frac{\omega }{\sqrt{1-k\tilde
r^{2}/a^{2}}}dt+\int k_{\tilde r}d\tilde r.
\end{equation}
Substituting the action into the Hamilton-Jacobi equation, one has
\begin{equation}
-\frac{\omega ^{2}}{1-k\tilde r^{2}/a^{2}}+\frac{2H\tilde r\omega
}{\sqrt{1-k\tilde r^{2}/a^{2}}} k_{\tilde r}+(1-\frac{\tilde
r^{2}}{\tilde r_{A}^{2}})k_{\tilde r}^{2}+m^{2}=0,
\end{equation}
which has solutions
\begin{equation}
\label{eq10}
 k_{\tilde r}=\frac{-H\tilde r\pm \sqrt{H^{2}\tilde
r^{2}+(1-\tilde r^{2}/\tilde r_{A}^2) [1-m^{2}(1-k\tilde
r^{2}/a^{2})/\omega ^{2}]}}{(1-\tilde r^{2}/\tilde
r_{A}^{2})\sqrt{ 1-k\tilde r^{2}/a^{2}}} \omega,
\end{equation}
where the plus/minus sign corresponds to an outgoing/incoming
mode. Now we consider an incoming mode since the observer is
inside the apparent horizon, like the case of particle tunneling
for the cosmological event horizon in de Sitter
space~\cite{Parikh}. It is obvious that the action $\mathbf{S}$
has a pole at the apparent horizon. Through a contour integral, we
obtain an imaginary part of the action
\begin{eqnarray}
\label{eq11}
 {\rm Im}  \mathbf{S} &=& -{\rm Im} \int \frac{H\tilde
r+ \sqrt{H^{2}\tilde r^{2}+(1-\tilde r^{2}/\tilde r_{A}^2)
[1-m^{2}(1-k\tilde r^{2}/a^{2})/\omega ^{2}]}}{(1-\tilde
r^{2}/\tilde r_{A}^{2})\sqrt{ 1-k\tilde r^{2}/a^{2}}} \omega
d\tilde
r \nonumber \\
&=& \pi \tilde r _{A}\omega.
\end{eqnarray}
Note from (\ref{eq10}) that in the coordinate (\ref{eq3}), there
is no contribution of an outgoing particle to the imaginary part
of the action, as in the case of an ingoing particle in the black
hole spacetime~\cite{Padd,PW}. In the WKB approximation, the
emission rate $\Gamma$ is the square of the tunneling amplitude
(here the particle tunnels from outside to inside the apparent
horizon)
\begin{equation}
\label{eq12}
 \Gamma \propto   \exp (-2 {\rm Im } \mathbf{S}).
\end{equation}
Combining (\ref{eq12}) with (\ref{eq11}), one can see clearly that
the emission rate can be cast in a form of thermal spectrum,
$\Gamma \sim \exp (-\omega/T)$,  with temperature
\begin{equation}
\label{eq13}
 T= \frac{1}{2\pi \tilde r_A}.
\end{equation}
Thus we have finished the proof that an observer inside the
apparent horizon will see a thermal spectrum with temperature
(\ref{eq13}) when particles tunnel from outside the apparent
horizon to inside the apparent horizon. This can be explained as
Hawking radiation of apparent horizon in the same spirit in the
tunneling approach proposed by Parikh and Wilczek that the Hawking
radiation of black hole is expressed as a tunneling phenomenon.
Furthermore, at this level of approximation, the mass of particle
does not enter the emission rate. This is just the remarkable
feature of thermal spectrum. In addition, let us stress here that
since the energy $\omega$ is measured by the observer with the
Kodama vector (\ref{eq5}), the thermal spectrum is therefore seen
by the same observer. That is to say, the Hawking temperature
(\ref{eq13}) is measured by the Kodama observer inside the
apparent horizon.

{\it Tunneling of Massless Particle:} Next we further show that
one can indeed assign the temperature (\ref{eq13}) to the apparent
horizon of the FRW universe by following the standard approach of
massless particle tunneling across the de Sitter
horizon~\cite{Parikh}. The basic idea is the same as the above, in
the semiclassical approximation (WKB approximation), the emission
rate can be related to the imaginary part of action of a system.
As in the case of de Sitter space~\cite{Parikh}, we will consider
the s-wave emission of massless particle. Higher partial wave
emission is suppressed by $\hbar$. In the s-wave approximation,
particles can be viewed as massless shells, and move along a
radial null geodesic.

The radial null geodesic for the metric (\ref{eq3}) obeys
\begin{equation}
\dot {\tilde r}=H \tilde r\pm \sqrt{H^2\tilde r^2+(1-\tilde
r^{2}/\tilde r_{A}^{2})},
\end{equation}
where the plus/minus sign corresponds to an outgoing/incoming null
geodesic. We consider an incoming geodesic since the particles
tunnel from outside to inside the apparent horizon. In addition,
let us note that we are only interested in the imaginary part of
action, therefore we need only to calculate the imaginary part
produced by the tunneling particles since remaining part is always
real. The imaginary part is produced by particles tunneling
through a barrier, the classically forbidden region. The imaginary
part can be obtained as follows.
\begin{equation}
\label{eq15}
 {\rm Im} \mathbf{S}= {\rm Im} \int^{\tilde
r_f}_{\tilde r_i}p_{\tilde r}d\tilde r = {\rm Im} \int^{\tilde
r_f}_{\tilde r_i} \int ^{p_{\tilde r}}_0 dp'_{\tilde r}d\tilde r,
\end{equation}
where $p_{\tilde r}$ is the radial momentum, $\tilde r_i$ is the
initial position, slightly outside the apparent horizon. And
$\tilde r_f$ is a classical turning point, there the semiclassical
trajectory can join onto a classical allowed motion. Furthermore,
using the Hamlitonian equation,
\begin{equation}
\dot {\tilde r} = \frac{\partial \tilde H}{\partial p_{\tilde
r}}=\left .\frac{d\tilde H}{dp_{\tilde r}}\right |_{\tilde r},
\end{equation}
where $\tilde H$ is the Hamlitonian of the particle, the generator
of the cosmic time $t$,  we can carry out the integration in
(\ref{eq15}) as follows,
\begin{eqnarray}
{\rm  Im }\mathbf{S} &=&  {\rm Im}
 \int_{\tilde r_i}^{\tilde r_f} d\tilde r \int d\tilde H \frac{1}{\dot
 {\tilde r}}
  \nonumber  \\
 &=&{\rm Im} \int_{\tilde r_i}^{\tilde r_f} d\tilde r \frac{\omega }
 {\dot {\tilde r}\sqrt{1-k\tilde r^2/a^2}}
 \nonumber \\
 &=& -\omega {\rm Im} \int_{\tilde r_i}^{\tilde r_f} \frac{d\tilde r}
  {\sqrt{1-k\tilde r^2/a^2}(\sqrt{
  H^2\tilde r^2+(1-\tilde r^{2}/\tilde r_{A}^{2})}-H\tilde r)}
  \nonumber \\
&=& \pi \tilde r_{A}\omega
\end{eqnarray}
Note that in de Sitter space case, the integration over the
Hamiltonian $\tilde H $ simply gives the energy $\omega$ of the
particle. In our case, it gives us $\omega/\sqrt{1-k\tilde
r^2/a^2}$ since our energy of the particle is measured by an
observer with the Kodama vector (\ref{eq7}). Thus using the
interpretation of emission rate (\ref{eq12}), once again, we
arrive at
\begin{equation}
T=\frac{\omega }{2 {\rm Im}\mathbf{S}}=\frac{1}{2\pi \tilde
r_{A}}.
\end{equation}
Thus we have shown again that the apparent horizon of FRW universe
has an associated temperature, $1/2\pi \tilde r_A$, like event
horizon of black hole.

{\it Conclusion and Discussions:} In summary, by using the
tunneling approach proposed by Parikh and Wilczek, we have
finished the proof that the apparent horizon of FRW universe has
indeed an associated Hawking temperature (\ref{eq1}), filling in
the gap existing in the literature. The Hawking temperature is
measured by an observer with the Kodama vector (\ref{eq5}) inside
the apparent horizon. With this, we can conclude that Hawking
radiation is not always associated with event horizon of
spacetime. That is to say, the existence of event horizon is not a
key cause of Hawking radiation, which was widely accepted before
in the community of black hole physics. In addition, some remarks
are in order. First, Hawking temperature is always related to
surface gravity of horizon as $T=|\kappa|/2\pi$, where $\kappa$ is
surface gravity of horizon. For the FRW universe, it is known that
the surface gravity of apparent horizon is~\cite{CK}: $\kappa
=-(1-\dot{\tilde r}_A/(2H\tilde r_A))/\tilde r_A$. In deriving
Friedmann equations by using $\delta Q= TdS$, a key point is to
calculate the amount of energy crossing the apparent horizon in an
infinitesimal time interval. During the infinitesimal time
interval, the radius of the apparent horizon is assumed to be
fixed, that is, $\dot{\tilde r}_A=0$. Thus one is led to $T=1/2\pi
\tilde r_A$. On the other hand, the tunneling process discussed in
the present paper is an instantaneous one, one has naturally
$\dot{\tilde r}_A=0$. Therefore we have the Hawking temperature
(\ref{eq1}) for the apparent horizon of FRW universe. In general,
one believes there is a well-known relation between Hawking
temperature and surface gravity for spacetime horizons,
$T=|\kappa|/2\pi$. Therefore, it is of great interest to see the
recovery of the relation by improving our discussions. Second, in
the present paper, since did not consider the back-reaction of
Hawking radiation, we therefore obtained an exact thermal spectrum
of apparent horizon radiation. Naturally, if the back-reaction is
taken into account, the radiation spectrum will deviate from the
thermal spectrum. Since the deviation from the thermal spectrum of
black hole is intensively discussed in the literature, we do not
repeat here. Third, our proof can be easily generalized to the
case of higher dimensional FRW universe with the same temperature
(\ref{eq1}). Finally we stress that although we get the Hawking
temperature in the coordinates (\ref{eq3}), one can reach the same
conclusion starting with the metric (\ref{eq2}) within the
Hamilton-Jacobi method.


\section*{Acknowledgments}
  This work was supported partially by
grants from NSFC, China (No. 10821504 and No. 10525060), and a
grant from the Chinese Academy of Sciences with No.KJCX3-SYW-N2.

\end{document}